 \documentclass[final,1p,times,authoryear]{elsarticle}
\usepackage{lipsum}
\usepackage{amssymb}
\usepackage{color}
\usepackage{soul}
\begin{document}

\begin{frontmatter}



\title{Multi-scaling of wholesale electricity prices}


\author{F. Caravelli$\ ^{1,2,3}$, J. Requeima$\ ^{1}$, C. Ududec$\ ^{1}$, A. Ashtari$\ ^{1}$, T. Di Matteo$\ ^{4}$ and T. Aste$\ ^{2,5}$}

\address{
$\ ^{1}$ Invenia Technical Computing, 135 Innovation Dr., Winnipeg, MB R3T 6A8, Canada\\
$\ ^{2}$Department of Computer Science, UCL,Gower Street, London, WC1 E6BT, UK\\
$\ ^{3}$London Institute of Mathematical Sciences, 35a South Street, London W1K 2XF, UK\\
$\ ^{4}$Department of Mathematics, King's College London, Strand, London WC2R 2L, UK \\
$\ ^{5}$Systemic Risk Centre, London School of Economics and Political
Sciences, London, WC2A2AE, UK}

\begin{abstract}
We empirically analyze the most volatile component of the electricity price time series from two North-American wholesale electricity markets.  We show that these time series exhibit fluctuations which are not described by a Brownian Motion, as they show multi-scaling, high Hurst exponents and sharp price movements. We use the generalized Hurst exponent (GHE, $H(q)$) to show that although these time-series have strong cyclical components, the fluctuations exhibit persistent behaviour, i.e., $H(q)>0.5$.  We investigate the effectiveness of the GHE as a predictive tool in a simple linear forecasting model, and study the forecast error as a function of $H(q)$, with $q=1$ and $q=2$.  Our results suggest that the GHE can be used as prediction tool for these time series when the Hurst exponent is dynamically evaluated  on rolling time windows of size $\approx 50 - 100$ hours. These results are also compared to the case in which the cyclical components have been subtracted from the time series, showing the importance of cyclicality in the prediction power of the Hurst exponent.
\end{abstract}


\end{frontmatter}

\section{Introduction}
Organized wholesale electricity markets are increasingly replacing vertically integrated systems around the world. 
These markets attempt to balance the objectives of short term stability and performance with economic competitiveness, efficiency, and long term planning (\cite{Hogan}, \cite{Liu1}, \cite{Loskow}, \cite{opti}).
Their structure is designed to take into account the unique properties of electricity among commodities: (i) it currently cannot be economically stored, so supply and demand must be continually balanced, (ii) injections into the transmission network flow according to Kirchhoff's laws, not from point to point, (iii) the transmission network has finite capacity and many operating constraints which limit the size of injections and withdrawals at specific locations, and (iv) the amount of power demanded is typically insensitive to price over short time periods.

In particular, these properties can lead to situations in which the network becomes segmented into effectively separate regions, meaning that some producers cannot competitively sell energy at certain locations.  This segmentation is usually called transmission congestion, and can lead to large inefficiencies and costs to consumers.
The standard approach to dealing with the above is to use a spatial and temporal pricing mechanism that explicitly accounts for the physical and operating constraints on the transmission network and generation units (\cite{Bohn84}).  
In this approach, prices at various locations in the transmission network, which are called Locational Marginal Prices (LMPs), are determined with the objective of maximizing social welfare while taking into account all physical and operating constraints, and compensating production and transmission providers for the marginal costs of their services.

These prices exhibit a very rich structure: they are high frequency, non-stationary, very seasonal, very volatile at times of peak demand and supply shortages, and fat tailed.  
From many points of view, such as financial risk management, market efficiency, mitigating market power, as well as short and long term planning, it is therefore important to understand the structure and predictability of these prices (\cite{Aggarwal}, \cite{Moest}, \cite{Benth}, \cite{Bottazzi}, \cite{Wang}, \cite{Weron}).

In this paper we show that the fluctuations in the most volatile component of LMPs exhibits multi-scaling behaviour and high generalized Hurst exponent values. In addition, we study the forecast error of simple linear forecasting models, and how this is related to the generalized Hurst exponent.  

In Section \ref{sec:data} we discuss the structure of the available data and its properties. In Section \ref{sec:analysis} we study the seasonality properties of the prices and the multi-scaling behaviour by using the generalized Hurst exponent approach. In Section \ref{sec:linmod} we connect the properties of the generalized Hurst exponent to the forecast error of a linear predictive model. We conclude in Section \ref{sec:conc}.

\section{Properties of the data}\label{sec:data}
 
The electricity markets in the United States have adopted what is known as a multi-settlement structure.
Market participants freely make bids and offers for electricity at different locations in the transmission network which are fed into a centrally organized dispatch and scheduling system administered by an independent authority, often called an independent system operator (ISO).
This bidding and dispatch is managed in a set of successive runs. 
The first run, performed a day in advance of an operating day, is called the Day-Ahead (DA) market. 
Using the DA bids, the ISO establishes a generation and load schedule and DA LMPs for each hour of the operating day, at each location (often called node) of the network.
The DA market exists to help participants and the ISO plan ahead, increase market liquidity, and mitigate market power (\cite{Loskow}).   
Then, as each hour of the operating day approaches, a Real-Time (RT) energy market is administered based on revised generation offers and load forecast. 
This process results in new hourly RT LMPs at each node.
RT LMPs are much more volatile than DA LMP's due to changes in fuel costs, unanticipated demand that must be met in real time by existing capacity, as well as planned and unplanned generation and transmission outages.

\begin{figure*}
  \centering
  \includegraphics[scale=0.4]{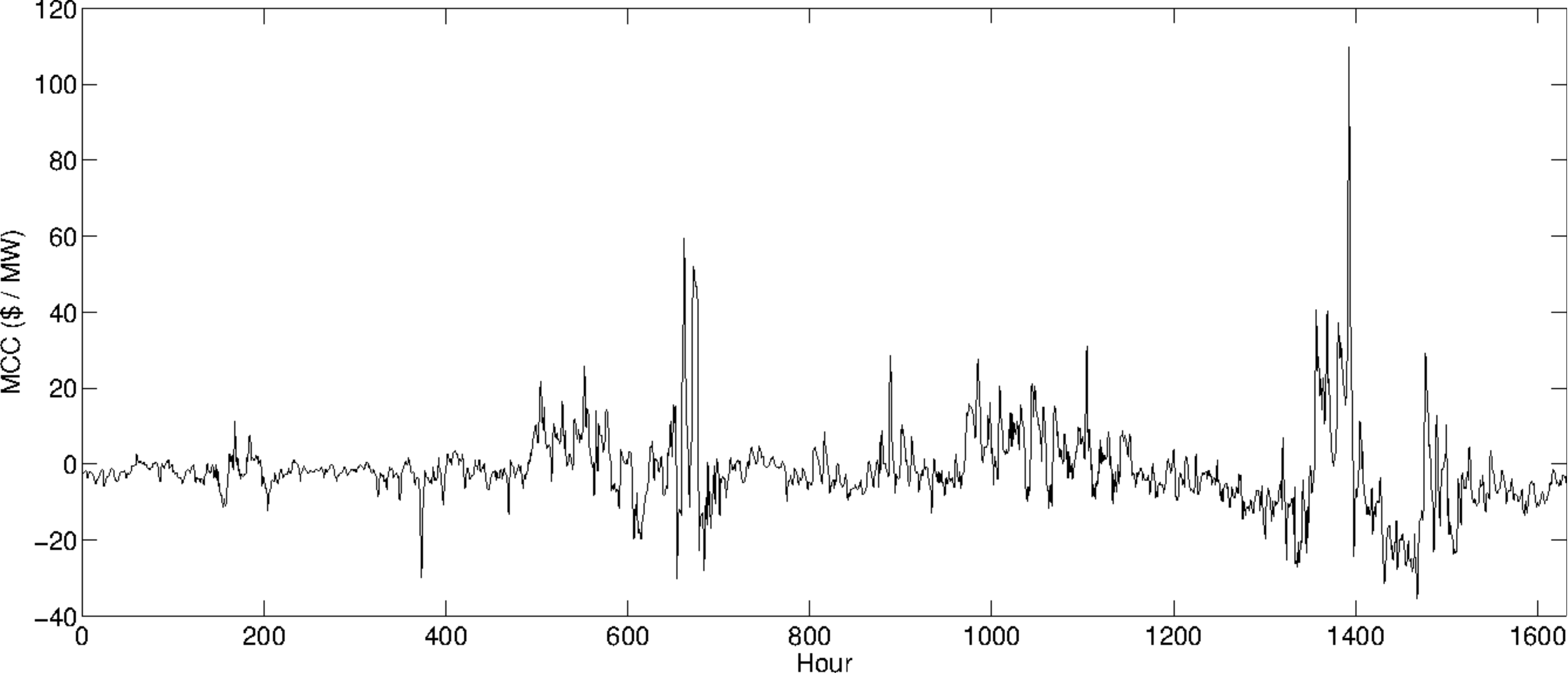}
  \caption{An example of the RT MCC component time series in MISO in the time period of 1632 hours starting on January 1, 2014. The price is the cost per Megawatt-hour of electricity at a particular node. We observe that the congestion component can be both positive and negative and that the price varies over many orders of magnitude. }
  \label{fig:example}
  \end{figure*}

In more technical terms, the LMP at some node is the marginal cost (expressed in \$/MWh) of supplying, at the lowest cost, the next increment of demand at that node, while taking into account supply and demand and the physical and operational constraints of the transmission network. 
For each node $n$ the DA or RT LMP at time $t$ can be split into three components (\cite{Liu1}):
\begin{equation}
LMP(n,t) = MEC(t) + MLC(n,t) + MCC(n,t),
\end{equation}
where MEC is the \textit{Marginal Electricity Component}, MLC is the \textit{Marginal Loss Component}, and MCC stands for \textit{Marginal Congestion Component}.  
The $MEC(t)$ component is the price of electricity at any given node if there is no congestion or loss to that node.  $MLC$ is the cost of transmission losses to a node, and is generally small in comparison with $MEC$.
The $MCC$ component is of particular interest to us: it is related to the congestion occurring in the transmission network and is the most volatile time series in RT and DA, as can be observed in Fig. \ref{fig:example} for a particular node in the MISO market.   

For our study we focus on the DA MCC component of price from two markets: the Midwest Independent System Operator (MISO) and Pennsylvania-New Jersey-Maryland Interconnection (PJM), for a period of time of 1632 hours, starting January 1, 2014 and ending March 9, 2014. There are 1287 nodes for PJM and 2568 nodes in MISO.

\section{Seasonality and multi-scaling}\label{sec:analysis}

Given the fact that LMPs are strongly influenced by the local and regional climates and supply and demand patterns, one could ask whether there are any cyclic patterns in LMPs, and if this influences the predictability of the time series (\cite{Brown}). 

\begin{figure*}
\centering
\includegraphics[scale=0.42]{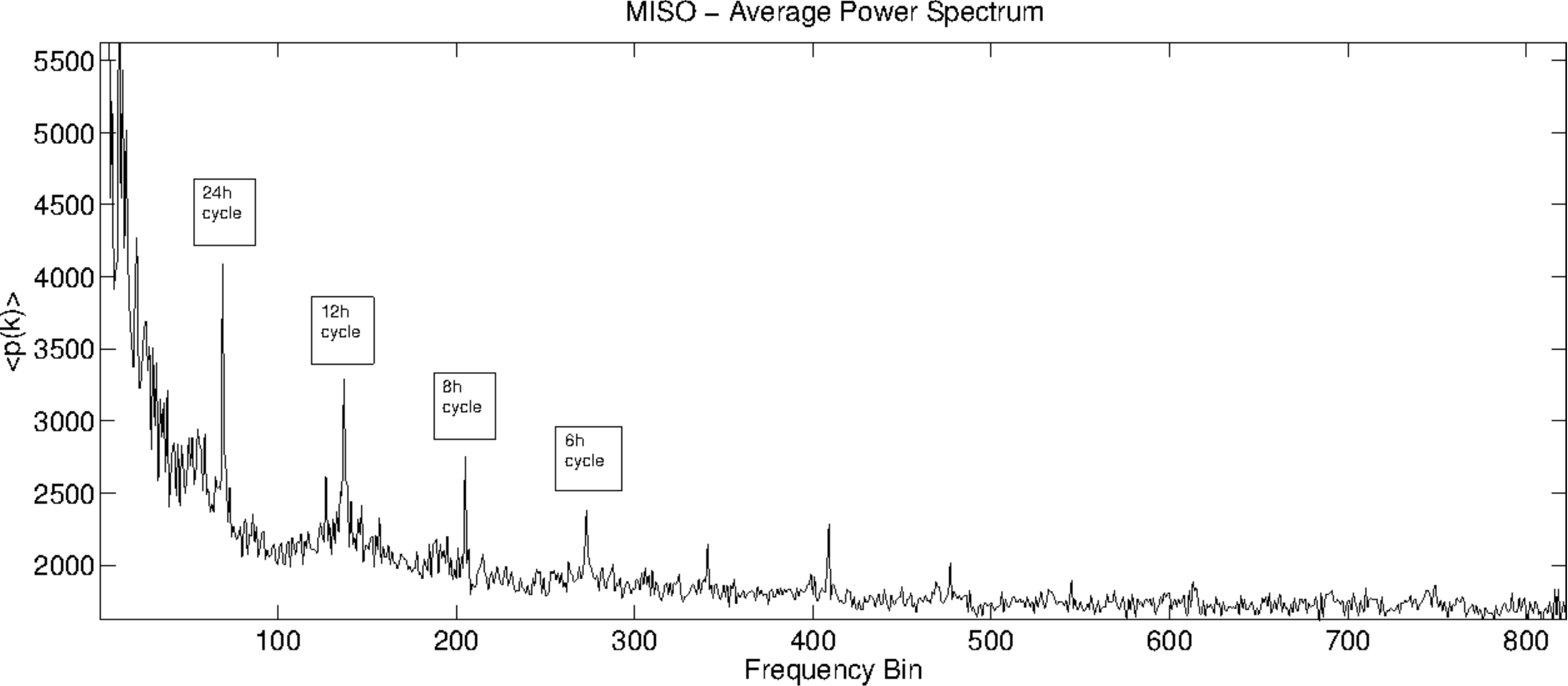}
\includegraphics[scale=0.42]{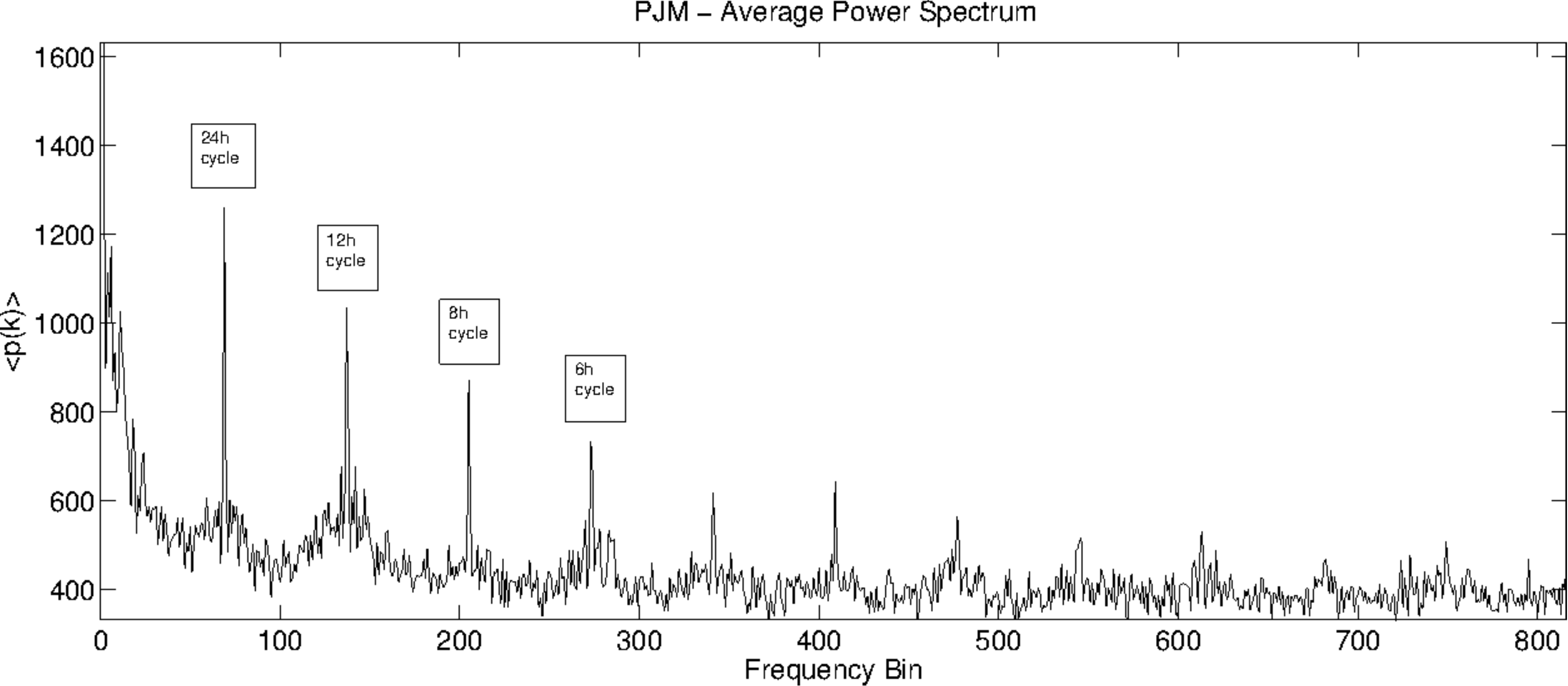}
\caption{Average power spectrum of MISO (top) and PJM (bottom) DA MCC prices, obtained through a Fast Fourier Transform with averaging over all the nodes in each market. The power spectrum has been calculated from a time series of 1632 hours with sample rate of one hour. }
\label{fig:PowerSpectrumFFT}
\end{figure*}

%\subsection{Frequency domain}
For this purpose, we study the discrete Fourier transform (computed with a Fast Fourier Transform (FFT) algorithm) of the time series of prices $S_i(t)$ at each node $i$, from the PJM and MISO markets.
In Fig. \ref{fig:PowerSpectrumFFT} we plot the power spectrum of each time series, defined as $p_i(k)=|\tilde S_i(k)|^2$, with $\tilde S_i(k)=FFT(S_i(t))$, evaluated over the whole time range and averaged over all nodes in the market.
The value in each bin $k$ is given by $\langle p(k) \rangle=\frac{1}{N} \sum_{i=1}^N p_i(k)$, where $N$ is the number of nodes in the market.
One can observe the presence of several peaks in the power spectrum, which one can identify  respectively as 24h, 12h and 8h cyclic patterns (see also \cite{Popova}, \cite{Weron}). 
These components have a larger size than the rest of the spectrum. 

%\begin{figure}
%\centering
%\includegraphics[scale=0.6]{SpectrumReduction.jpg}
%\includegraphics[scale=0.4]{FourierReduction.jpg}
%\end{figure}

%\begin{figure}
%\centering
%\includegraphics[scale=0.37]{Wavelet1.jpg} \includegraphics[scale=0.37]{Wavelet2.jpg}
%\label{fig:stationarity}
%\end{figure}

%\subsection{Time domain}
In order to investigate the properties of the fluctuations of the prices, we next analyze the scaling properties using a Hurst exponent approach (\cite{Hurst}, \cite{Mandelbrot}). The Hurst exponent and its generalizations have been used to study financial market behaviours (for an introduction see \cite{Beran} and \cite{Baillie}). Its importance has been stressed in the last decade in a series of papers 
(\cite{DM2}, \cite{Bouchaud},  \cite{DM1}, \cite{Lillo1}, \cite{Farmer1},  \cite{Bartolozzi1},  \cite{Barunik},  \cite{Morales2}, \cite{Barunik2}, \cite{Morales1}, \cite{Multifr}, \cite{Duan1}), and it has been used as a tool for quantifying the different degree of development and efficiency of various financial markets (\cite{DiMatteoDev}). Multi-scaling has been investigated also through de-trended fluctuation analysis (see for instance \cite{Multifr}, \cite{PredEM} and references therein). In this paper we use the Generalized Hurst Exponent (GHE) approach. 

\begin{figure}
\centering
\includegraphics[scale=0.36]{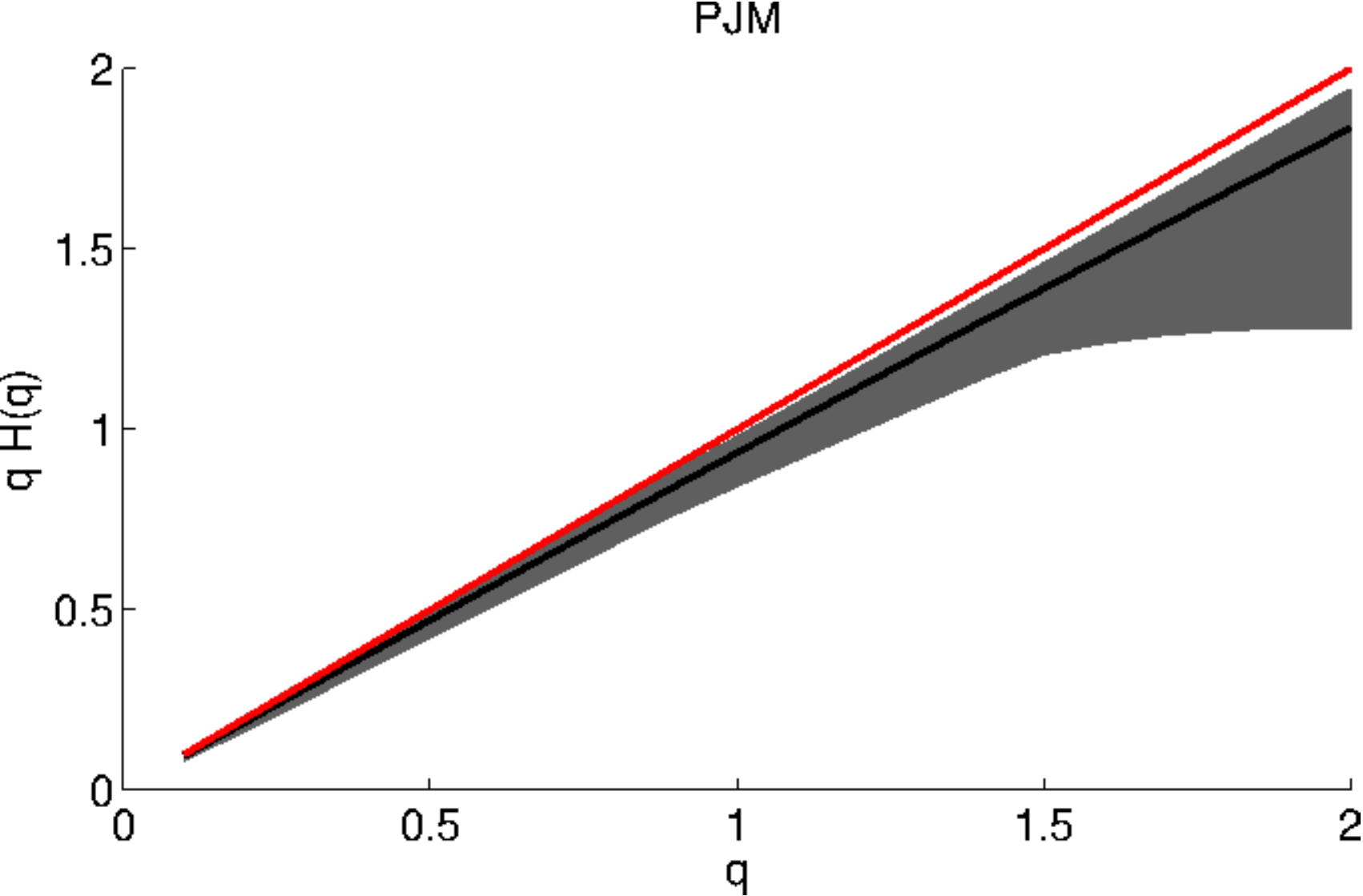} \includegraphics[scale=0.36]{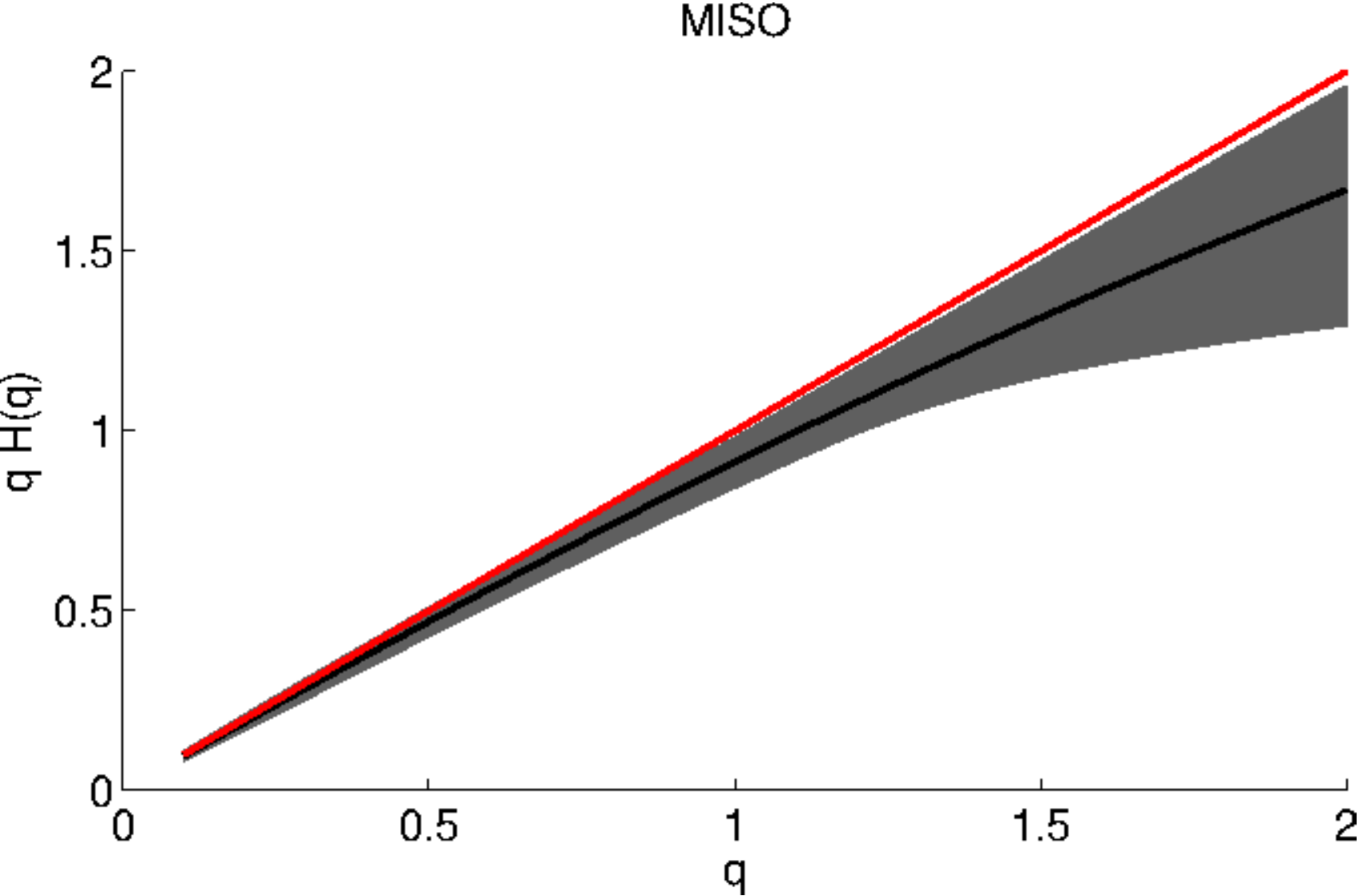}
\caption{Analysis of the exponent $q H(q)$ in Eq. (\ref{eq:hexp}) for PJM (left) and MISO (right). The red line represents the $y=x$ line, and the black line is the average over the market of the function $q H_i(q)$. The shaded area represents the minimum and the maximum over the market. This figure shows that in MISO market the multi-fractal behaviour is more pronounced than in PJM.}
\label{fig:multifract}
\end{figure}

The GHE is a tool used to study the statistical and scaling properties of time series. The scaling is characterized by an exponent $H(q)$ which is commonly associated with the long-term statistical dependence of the series. 
It is defined from the scaling of the q-th order moments of the distribution of increments  (\cite{DMRev}, \cite{DM2}) 
\begin{equation}
K_q (\tau)= \frac{\langle |X(t+\tau)-X(t)|^q\rangle}{\langle |X(t)|^q\rangle}\approx \Big(\frac{\tau}{\nu}\Big)^{q\  H(q)}
\label{eq:hexp}
\end{equation}
where the time interval $\tau$ can vary between $\nu$ and $\tau_{max}$, and $X(t)=cumsum(S(t))$ is the cumulative sum of the time series of interest $S(t)$, and $\nu$ is the sampling rate. 

% The Generalized Hurst exponent $H(q)$ can be defined from the scaling behaviour of $K_q(\tau)$  (\cite{DMRev}, \cite{DM2}) as in Eq. (\ref{eq:hexp}). 

Within this framework, two kinds of processes can be distinguished: (i) a process where $H(q)=H$ is constant and independent of $q$; and (ii) a process with $H(q)$ not constant. The first case is characteristic of uni-scaling or uni-fractal processes and the scaling behaviour is determined from a unique constant $H$ that coincides with the Hurst coefficient or the self-affine index. This is the case for self-affine processes where $q\ H(q)$ is linear ($H(q)=H$) and fully determined by its index $H$. In the second case, when $H(q)$ depends on $q$, the process is commonly called multi-scaling (or multi-fractal) and different exponents characterize the scaling of different $q$-moments of the distribution.
The Hurst exponent is related with the the scaling behavior observed in power spectra $\tilde X(k)\approx k^{-\beta}$ by $\beta=1+ 2 H(2)$ (\cite{DM2}).

It was noticed in (\cite{DM2}) that the Hurst exponents evaluated using Eq. (\ref{eq:hexp}) do not strongly depend on the choice of detrending procedure used on the time series or on $\tau$ if this is taken sufficiently large. 
For the present paper we have verified that using detrendings over time windows of size $\delta t \in[6, 12]$ hours,  the changes of the values of the Hurst exponents were consistently below 10\%.

%Different colors correspond to different $q$, and we observe that for all the various $q$ and market nodes, $\epsilon_i(q)\leq 10\%$, already for small $\delta h\approx 50$ hours. 
%\textcolor{red}{expand on using the cumsum and not the actual time series-
We analyze the values of the Hurst exponents for the PJM and MISO cumulative time series i.e., we evaluate the Hurst exponent on $X(t)=cumsum(S(t))$, where $S(t)$ are the DA MCC prices.
This approach is justified by the fact that in the Day-Ahead market, participants place bids one day ahead for each hour, and thus the cumulative sum of the time series is the real return as a function of time.
In Fig. \ref{fig:multifract} we show the multi-scaling behavior in these markets by plotting the average quantity $q H(q)$ in Eq. (\ref{eq:hexp}) (dark line) together with the shaded area representing the standard deviation, and observing that these curves are below their linear trend. Fig. \ref{fig:multifract} shows that prices in both markets exhibit strong multi-fractal behaviour, although this phenomenon is more pronounced in MISO. These results are in line with (\cite{Bottazzi}) for the case of North-European electricity markets, and can be attributed to the fact that electricity is not economically storable.
 
\begin{figure}
\centering
\includegraphics[scale=0.36]{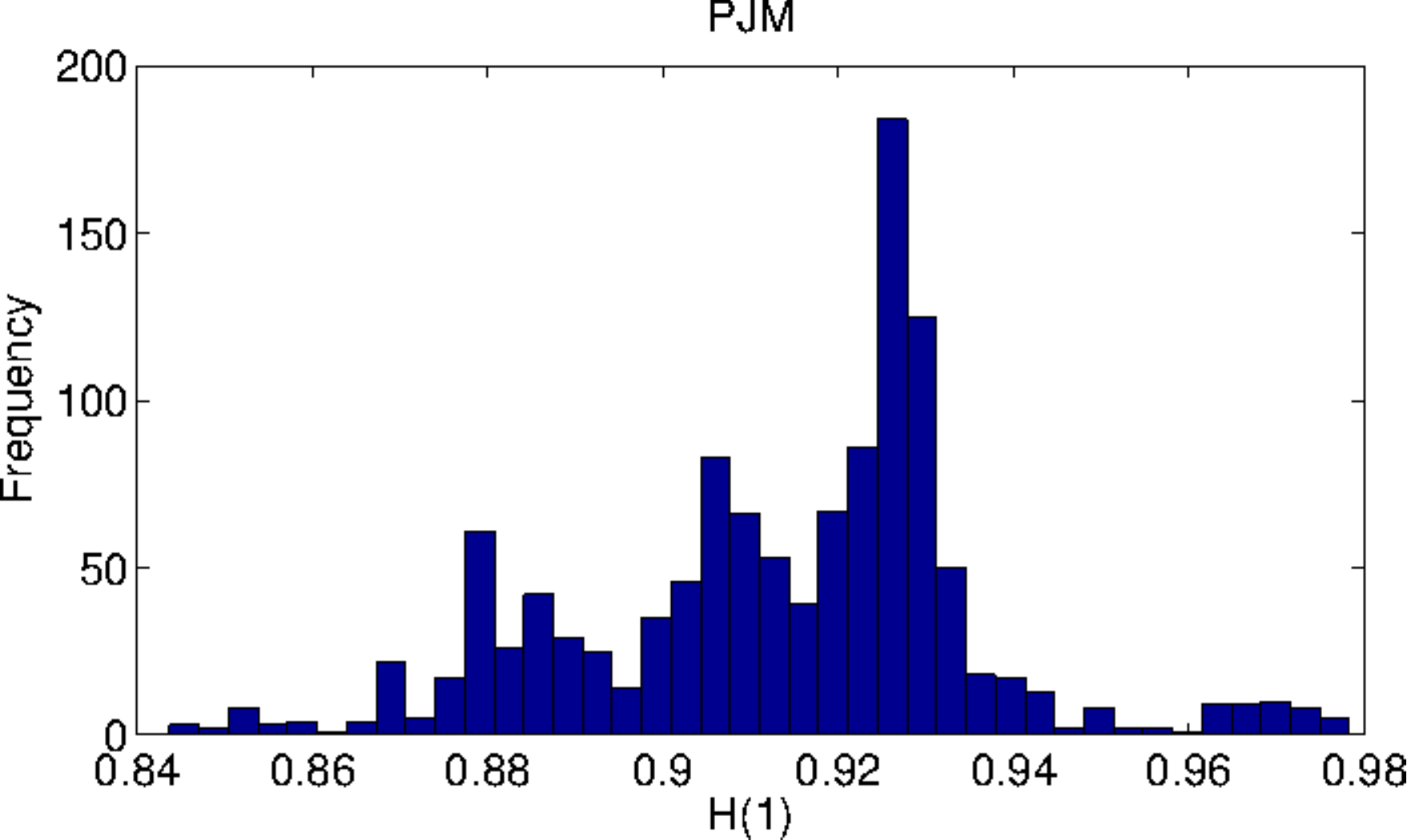} \includegraphics[scale=0.36]{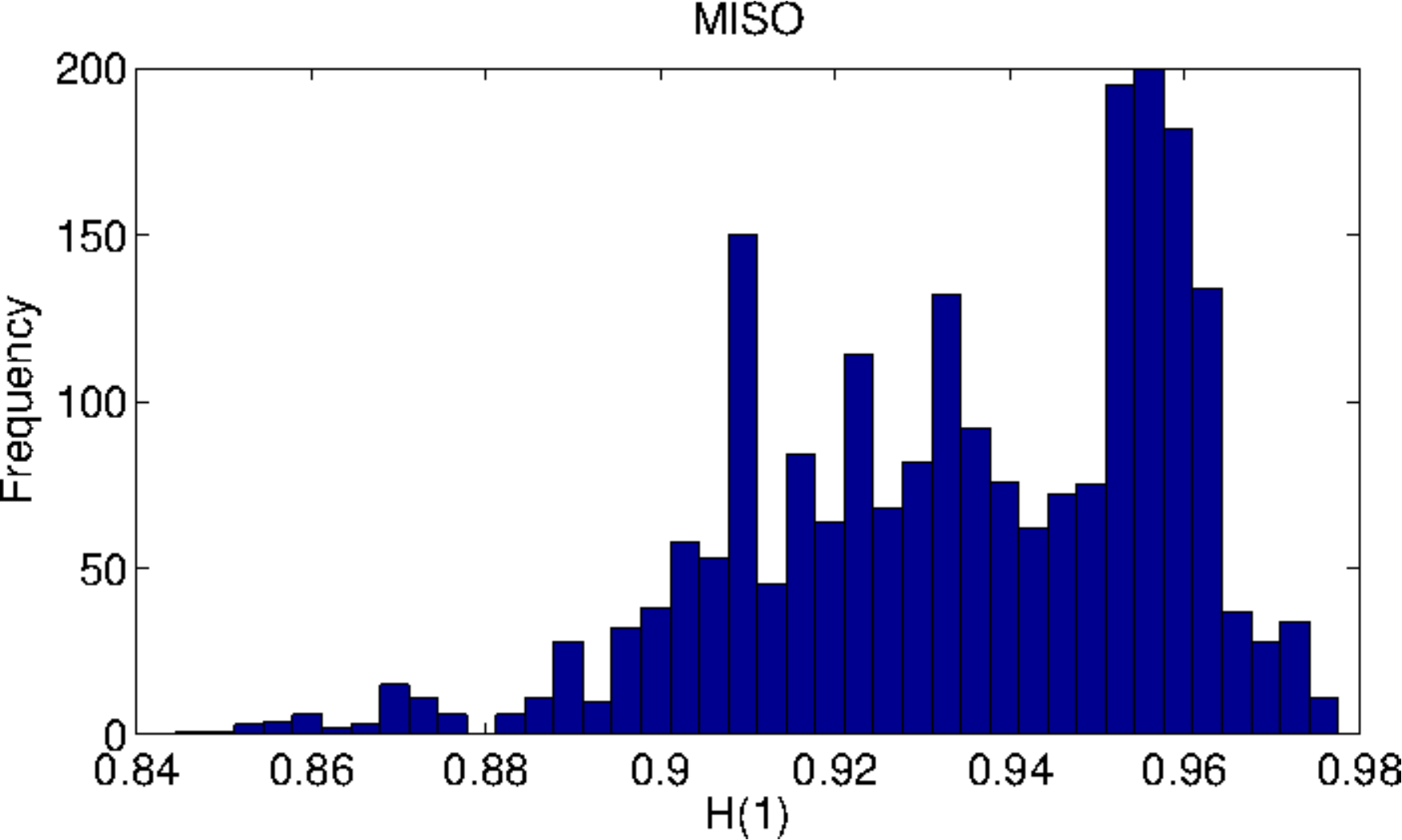}
\caption{Descriptive statistics of the Hurst exponent $H(1)$, evaluated on the whole time series for each node, for PJM (left) and MISO (right), and for $\tau_{max}=19$ days and averaged over the whole time series.}
\label{fig:HurstHist}
\end{figure}

\begin{figure}
\centering
\includegraphics[scale=0.36]{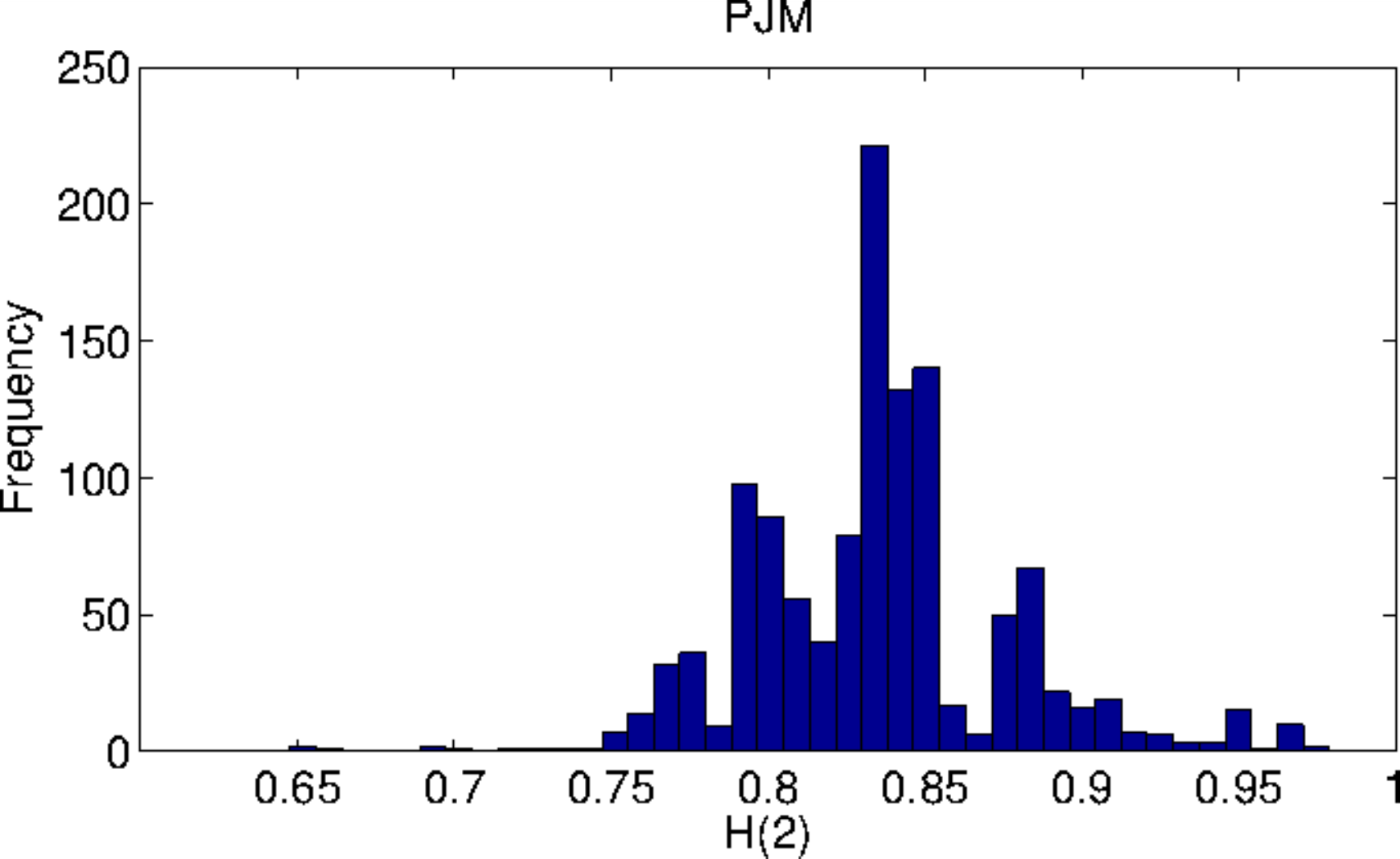} \includegraphics[scale=0.36]{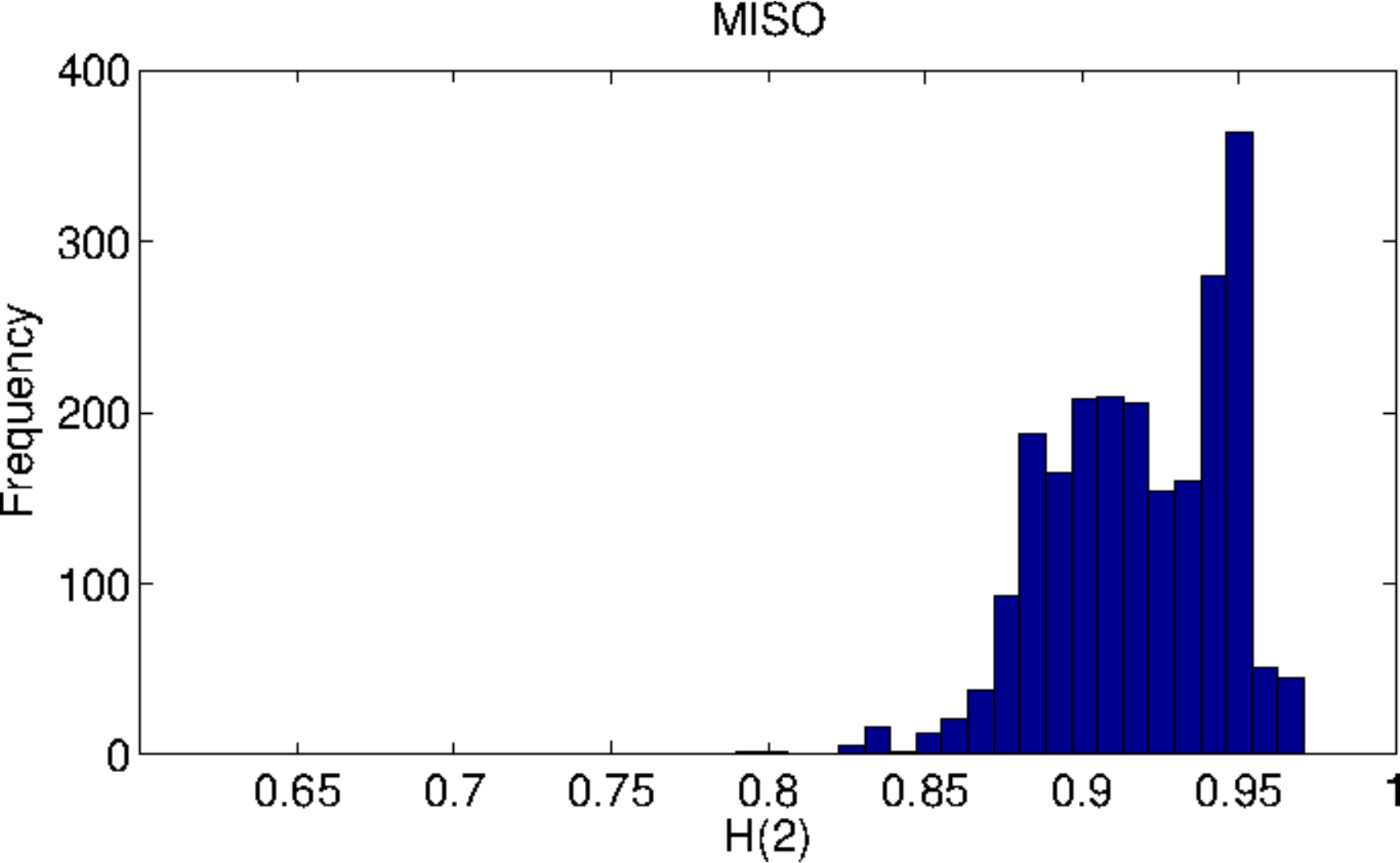}
\caption{Descriptive statistics of the Hurst exponent $H(2)$, evaluated on the whole time series for each node, for PJM (left) and MISO (right), and for $\tau_{max}=19$ days.}
\label{fig:HurstHist2}
\end{figure}

A snapshot of the distribution of the Hurst exponent $H(1)$ is shown in Fig. \ref{fig:HurstHist}, where we plot a histogram of $H(1)$ evaluated over the whole time series ($\delta h\approx1600$). We observe that $H(1)\in [0.84,0.98]$ in PJM and $H(1) \in [0.85,0.99]$ in MISO, strongly deviating from the expected value of a Brownian motion, $H=0.5$.  Comparing to $H(q)$ for $q=2$, in Fig. \ref{fig:HurstHist2} we plot the histogram of  $H(2)$ evaluated on the whole time series for each market, observing that $H(2)\in[0.65,0.95]$ in PJM, and $H(2)\in[0.8,0.96]$ in MISO. This shows that the Hurst exponent is indeed not constant with $q$, and in particular in the case of PJM, $H(2)$ is clustered around lower values than $H(1)$.  For both $H(1)$ and $H(2)$, the generalized Hurst exponents diverge dramatically from the Brownian motion value $H=0.5$.%\textcolor{red}{TIZIANA: Possiamo anche stressate il punto che gli H sono diversi da 0.5 a cui sembra che non diamo molto peso in questa versione del paper.} \textcolor{green}{This analysis suggests not only that the of Day-Ahead markets in MISO and PJM are multi-fractals, but that these have values of the exponent which is clustered around large values $H(q)>0.8$ both for $q=1$ and $q=2$.} 

%\textcolor{red}{TIZIANA: Non e? chiaro se fai un GHE pesato, dinamico o altro (figs 6 e 7). } 
%In order to picture how Hurst exponents fluctuate, we use a short interval $\delta h$ and plot the average Hurst exponents as a function of time, for both markets PJM and MISO. \textcolor{green}{
%%%%%%%%%
Let us now introduce an extra parameter:  the size, $\delta h$, of the moving windows on which we calculate the generalized Hurst exponents dynamically.
We observed that the Hurst exponents computed on short ($\delta h=50$ hours) and long ($\delta h\approx 1600$ hours) time windows have comparable results.
The dependence and importance of the choice of $\delta h$ for forecasting will be addressed specifically in the next section. 

\begin{figure*}
\centering
\includegraphics[scale=0.44]{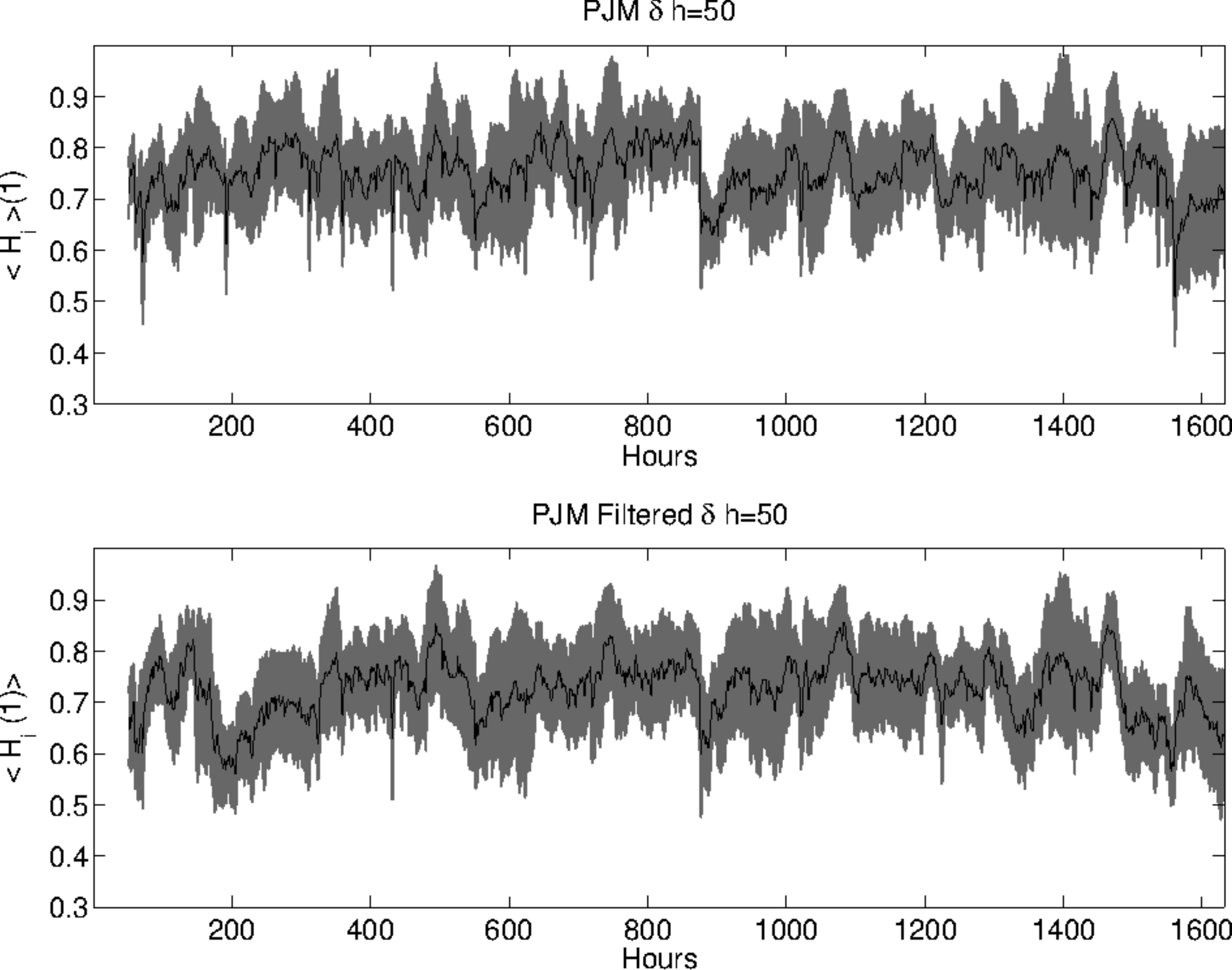}\ \\\ \\
\includegraphics[scale=0.44]{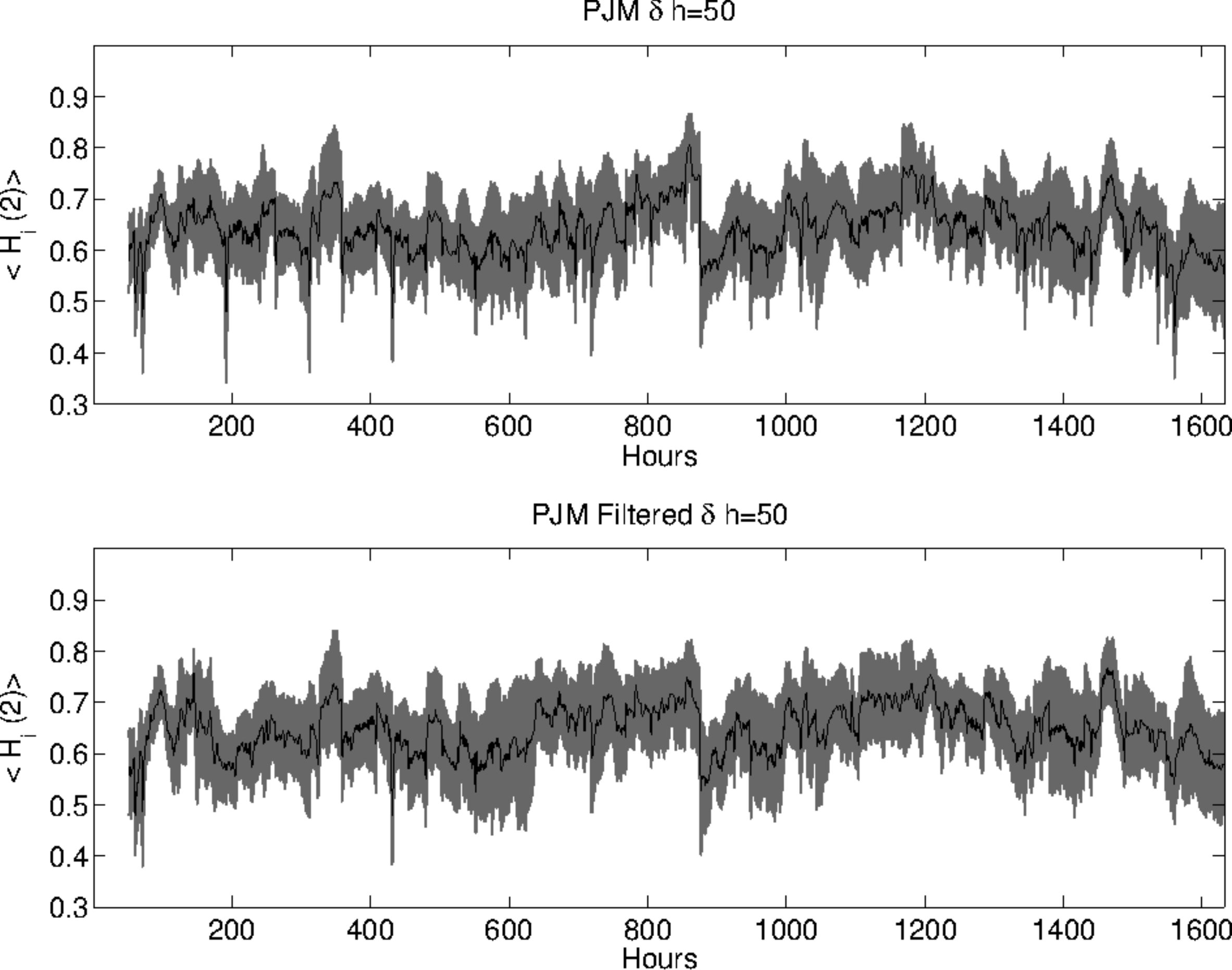}
\caption{Average H(1) and H(2) exponents for PJM evaluated dynamically, with and without filtering,  on a moving time window of  $\delta h=50$ hours. The dark line represents the average value across the market, meanwhile the shadowed area represents the standard deviation. We observe that for both filtered and unfiltered time series the average value of $H(1)$ oscillates around high values of the Hurst ($ \approx 0.5 - 0.9$), although some nodes occasionally have values below $H=0.5$, thus exhibiting also anti-persistent behavior.  $H(2)$ is consistently lower than $H(1)$, with $H(2)\approx 0.45 - 0.82$.  In general both exponents differ strongly from the Brownian motion expected value of $H=0.5$. We note that due to the short time window on which the Hurst is calculated here, these values are consistently lower than the one evaluated on the whole time series, as in Fig. \ref{fig:HurstHist}.  In addition, many of the spikes observed in the unfiltered signal disappear or are attenuated in the filtered case.}
\label{fig:HurstAveragePJM}
\end{figure*}

\begin{figure*}
\centering
\includegraphics[scale=0.44]{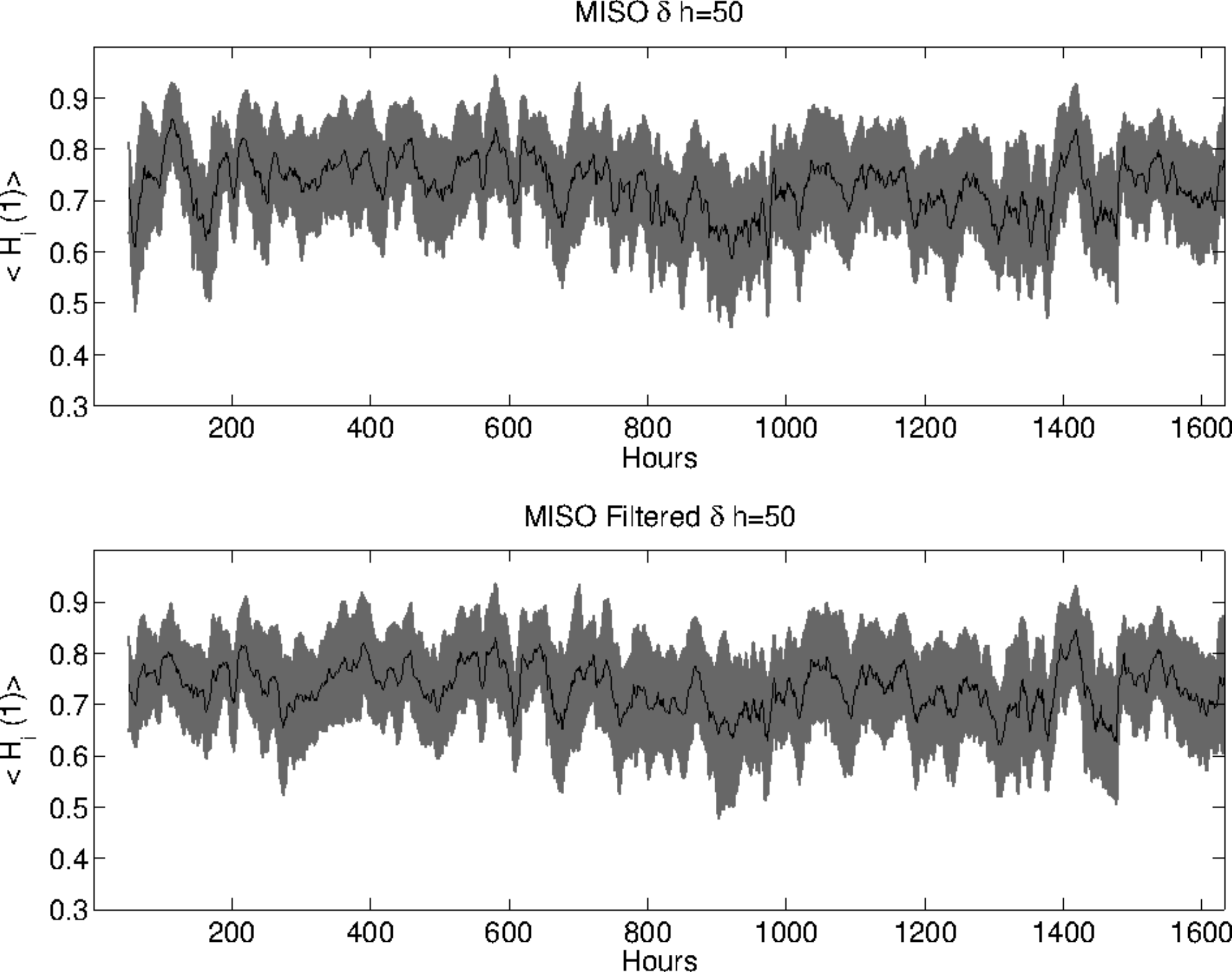}\ \\\ \\
\includegraphics[scale=0.44]{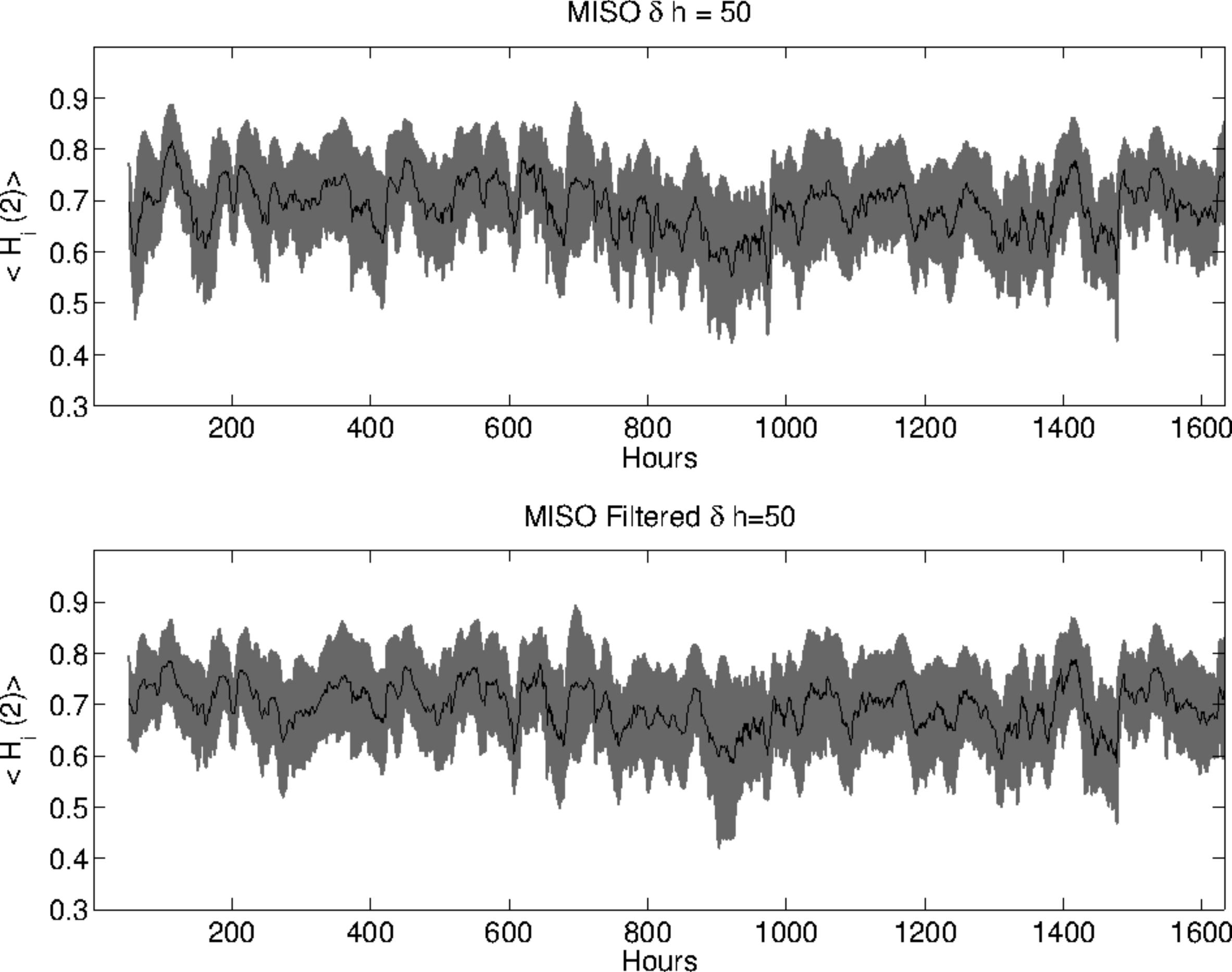}
\caption{Average H(1) and H(2) exponents for  MISO, with and without filtering, calculated on a moving time window of $\delta h=50$ hours, as in Fig. \ref{fig:HurstAveragePJM}. The dark line represents the average value across the market, and the shadowed area represents the standard deviation. We observe that for both markets the average values of $H(1)$ and $H(2)$ oscillate around high values, $H(1)\approx 0.5 - 0.9$ and $H(2)\approx 0.5 - 0.8$, and $H(2)$ has a systematically lower value than $H(1)$. Finally, the filtered signal shows less cyclicality and spiky behavior.}
\label{fig:HurstAverageMISO}
\end{figure*}

To begin with, we  evaluate the Hurst exponents dynamically in a moving window of length $\delta h=50$.  In Fig. \ref{fig:HurstAveragePJM} and Fig. \ref{fig:HurstAverageMISO}, the black line is the average generalized Hurst exponent, for $H(1)$ and $H(2)$ evaluated on the real and filtered signal.
In the latter case, we consider the signal with the frequencies which dominate the power spectrum subtracted (in particular, the 6h, 8h, 12h and 24h components). 
The average is taken over the nodes $i$ in the market, i.e., $\langle H_i (q)\rangle=\frac{1}{N} \sum_{i=1}^N H_i (q)$.  See  (\cite{Morales1}) for a similar analysis performed on stock market data.
The use of a short time window with $\delta h=50$ hours will be further motivated in the next section, where we will suggest the use of this training window to minimize forecast errors.
It is also interesting to note that even if evaluated on such a short window, the values of these exponents appear to be in the same ranges as those evaluated on the whole time series. 

For both PJM and MISO, the Hurst exponent is subject to rather strong fluctuations and spiky behaviour. We note that the latter is rather attenuated when we consider the signal from which the cyclic components have been removed, but still the Hurst exponents oscillate in the range $H(1) \approx  0.6 - 0.9$ and $H(2) \approx 0.5 - 0.85$ for both markets, which confirms the persistence of the fluctuations. Even after the signal has been filtered, we note that some abrupt transitions occur for both $H(1)$ and $H(2)$. Further, in Fig. \ref{fig:HurstAveragePJM} we show that there are changes in $H(1)$ and $H(2)$ which are coherent across the entire market. In the case of some of these sharp movements, we could identify specific weather events which could have potentially triggered this coherent behaviour. For instance, in the case of PJM, the sharp transition occurring at  $t\approx 880$h corresponds to a severe snowfall occurring in Pennsylvania on the 5th of February 2014. We thus argue that the behaviour of the fluctuations can be interestingly connected to specific exogenous factors.

% In order to visualize the multifractality, in Fig. \ref{fig:Multifractality} we plot the function $f_i(q)=q H_i(q)$ evaluated numerically for each node, in both markets, PJM and MISO. If the process underlying the fluctuations in the time series is a regular fractal, then one observes $f_i(q)$ being a linear function. If otherwise the process is multifractal, $f_i(q)$ is not linear, and indeed one observes that higher moments of the time series behave as $\langle |S(t+W)-S(t)|^q\rangle\approx W^{-q H(q)}$, hint that the generalized Hurst moment is not a constant. 
%In addition to this analysis, we evaluated the other exponents $H(q)$ and observed that, on average, $H(2)$ is lower that $H(1)$ in both markets. %Processes of this type are for instance Levy flights. 
%We observe that, while most of the nodes, in both markets, have a fairly linear behavior, and thus are well described by a fractal process, many nodes do indeed have a Hurst exponent which is different at different time scales.

%\begin{figure}
%\centering
%\includegraphics[scale=0.4]{PJMMultifractality.jpg}
%\includegraphics[scale=0.4]{MisoMultiFractality.jpg}
%\caption{Multifractality analisys, node by node, of PJM and MISO markets, calculated from the the whole time series. We plot $y_i(q)=q H_i(q)$ in the domain $q\in [0.1,2]$for every node $i$ in both markets. We have estimated the sensitivity on the evaluation of the Hurst value $H(q)$, which is within $10\%$ for all the values of $q$ considered. All the curves $q H_i(q)$ are sublinear for every node in both markets.  We observe different behavior, node by nodes, in both market, and in particular some nodes exhibit fluctuations which are rather fat tailed with a tail exponent which can be graphically extrapolated to a value which is clearly less than $2$. }
%\label{fig:Multifractality}
%\end{figure}

\section{Generalized Hurst exponent and trends} \label{sec:linmod}

Several studies have suggested that the Hurst exponent might be a measure of predictability of time series in stock markets (for instance \cite{Lillo1}, \cite{Farmer1}) and connected it to the efficient market hypothesis. It has been studied in detail in different models of stock markets as a measure of returns predictability in (\cite{Duan1}) and (\cite{Mitra}), and as a predictor of forecast quality for neural networks models in (\cite{Qian1}).
The relation between the Hurst exponent and the efficient market hypothesis has also been studied in (\cite{Eom1}, \cite{Eom2}). 
The predictability of Canadian electricity market prices using detrendend fluctuations is analysed in (\cite{PredEM}). 
There is also a large literature on the predictability of electricity prices, for examples see \cite{Aggarwal}, \cite{Moest}, \cite{Benth}, \cite{Wang}. 
Our focus will be on applying the Generalized Hurst exponent method for forecasting purposes. In order to do so, we introduce a simple linear regression model and measure the dependence of the forecast error on the Generalized Hurst exponent evaluated dynamically for each time series.

\subsection{The model}
Consider the simplest forecast method, a linear regression model.  In particular, for each time-series $S_i$, we extrapolate the next $p$-points by means of a linear regression performed on the previous $\delta h$ points: 
%\begin{equation}
$ \hat S_i(t+p)= \beta_1 p+\beta_0 $,
%\label{eq:pred}
%\end{equation}
%and where one estimates the parameters (or features) $\beta_1$  by minimization of the functional $\epsilon=\sum_{\xi=t-\delta h}^t (S_i(\xi)-\hat S_i(\xi))^2$ between the estimated values $\hat S_i(t)$ and the time series $S_i(t)$, 
%meanwhile constraining the intercept to match the last point on the interval.
with the intercept fixed at the last observation $\beta_0 = S(t)$.
It is well know that linear regression suffers from various problems, since it relies on the normal distribution of errors, does not take into account heteroscedasticity, and does not capture non-linear or chaotic patterns. Our goal here is to test the relationship between the Hurst exponent and the persistence of a trend as measured by a linear model.
%prediction capability with a simple model. %In general, linear regressions are  unstable when independent variables are highly correlated. Notwithstanding these features, we use linear regressions for their simplicity.
%where $\vec S_i=[S_i(t-\delta h),S_i(t-\delta h+1),\cdots, S_i(t)]^T$. Predicted values ahead of $p$ hours are given by $\hat S(t+p)=\beta_1 p+ \beta_0$. In general, one has that $\beta_0$ is simply the very last datapoint, $S_i(t)$, meanwhile $\beta_2$ is proportional to the autocorrelation of the time series.
These regressions are performed for each node in the market independently.  For each prediction, one can associate an error and two independent datapoints given by the values of the Generalized Hurst, $H_i(1)$ and $H_i(2)$ evaluated over the training window.  More precisely, we will study the relationship between forecast error and these Hurst exponents.
%} As an example of a series of predictions for both methods at different time points of a sinusoidal function is provided in Fig. \ref{fig:Spaghetti}.}
%\begin{figure}
%\centering
%\includegraphics[scale=0.4]{Spaghetti.jpg}
%\includegraphics[scale=0.33]{Regression.pdf} \includegraphics[scale=0.33]{ReitRegression.pdf} 
%\caption{Examples of the regression (left) and reiterated regression models (right) applied to a periodical time time series. At each time step, the linear predictor is a deterministic AR(1) regressor; in the regression case, the forecast is simply determined by the straight line, meanwhile in the reiterated case the new point is included into the training window and the regression performed again. }
%\label{fig:Spaghetti}
%\end{figure}

\subsection{Results}

%Thus, at each time step, one has effectively to estimate two parameters $a_{n,i}$ and $b_{n,i}$, which form the linear function
% \begin{equation}
% \tilde y_i(t)=a_{t,i} \tilde y_i(t-1)+b_{t,i},
% \end{equation}
% where the coefficients $a_{t,i}$ and $b_{t,i}$ are evaluated on a time window $\delta h$ on the values of the time series $[y_i (t-\delta h-1),\cdots,y_{i} (t-1)]$. For each prediction made at time $n$, we consider the predicted values of the time series at subsequent times as $\tilde y_t(t+n)$.

 %where the average is performed over all the data points which are available at the time $n$. If the predictor has $n$ predicted points in the future, in general, there will be $n$ generated different curves which pass from that point. We are then interested in the relation between $H(1)$, evaluated in a time window which overlaps with the data available to the linear predictor to estimate the parameters $a_{t,i}$ and $b_{t,i}$, and the average prediction error. We choose a prediction window of 24 hours. We then estimate the Hurst coefficient within the training window of length $\delta h$, and search for a correlation between the Hurst and the error.
 
\begin{figure}
\centering
\includegraphics[scale=0.28]{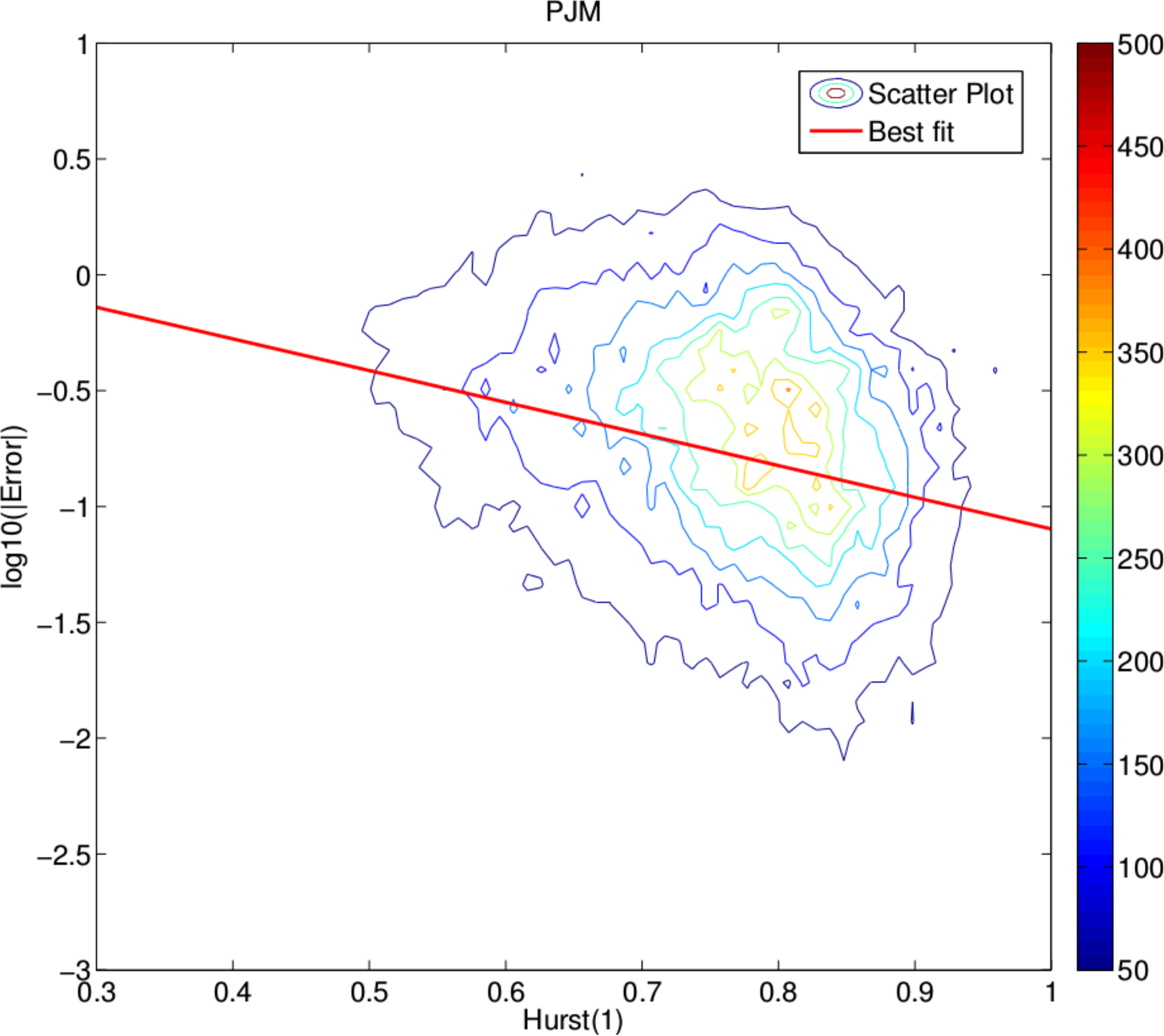} \includegraphics[scale=0.28]{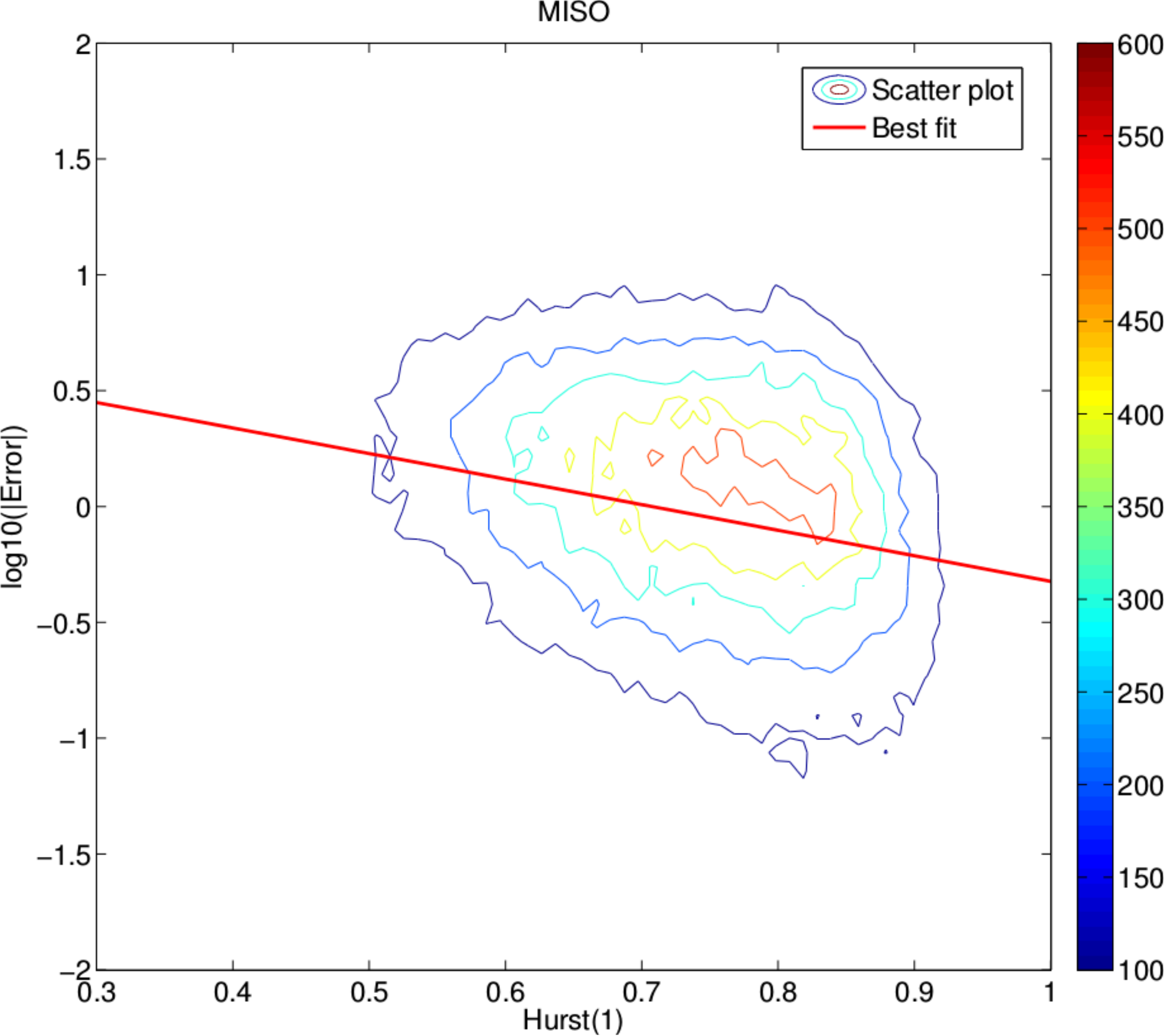}\\
\includegraphics[scale=0.28]{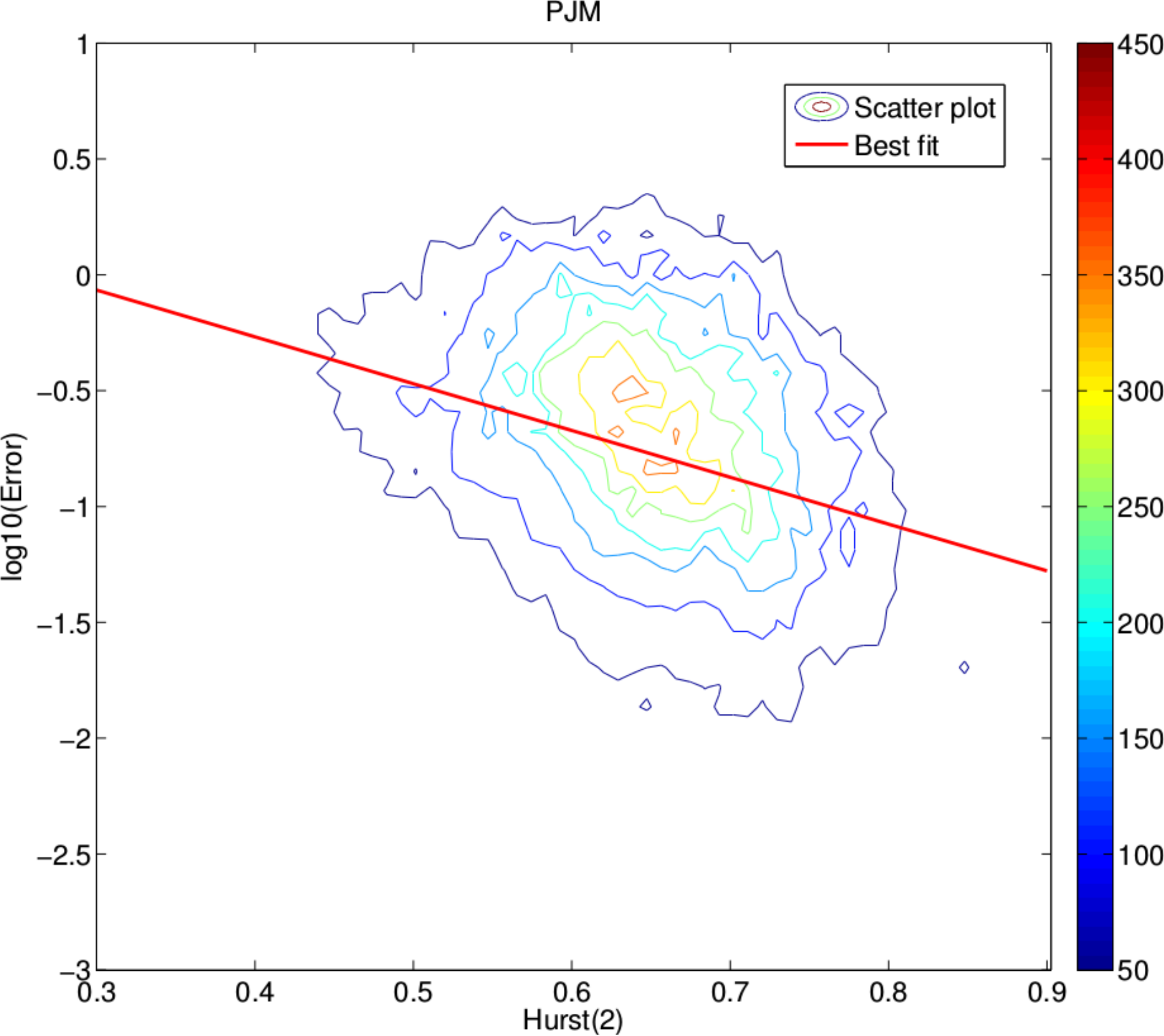} \includegraphics[scale=0.28]{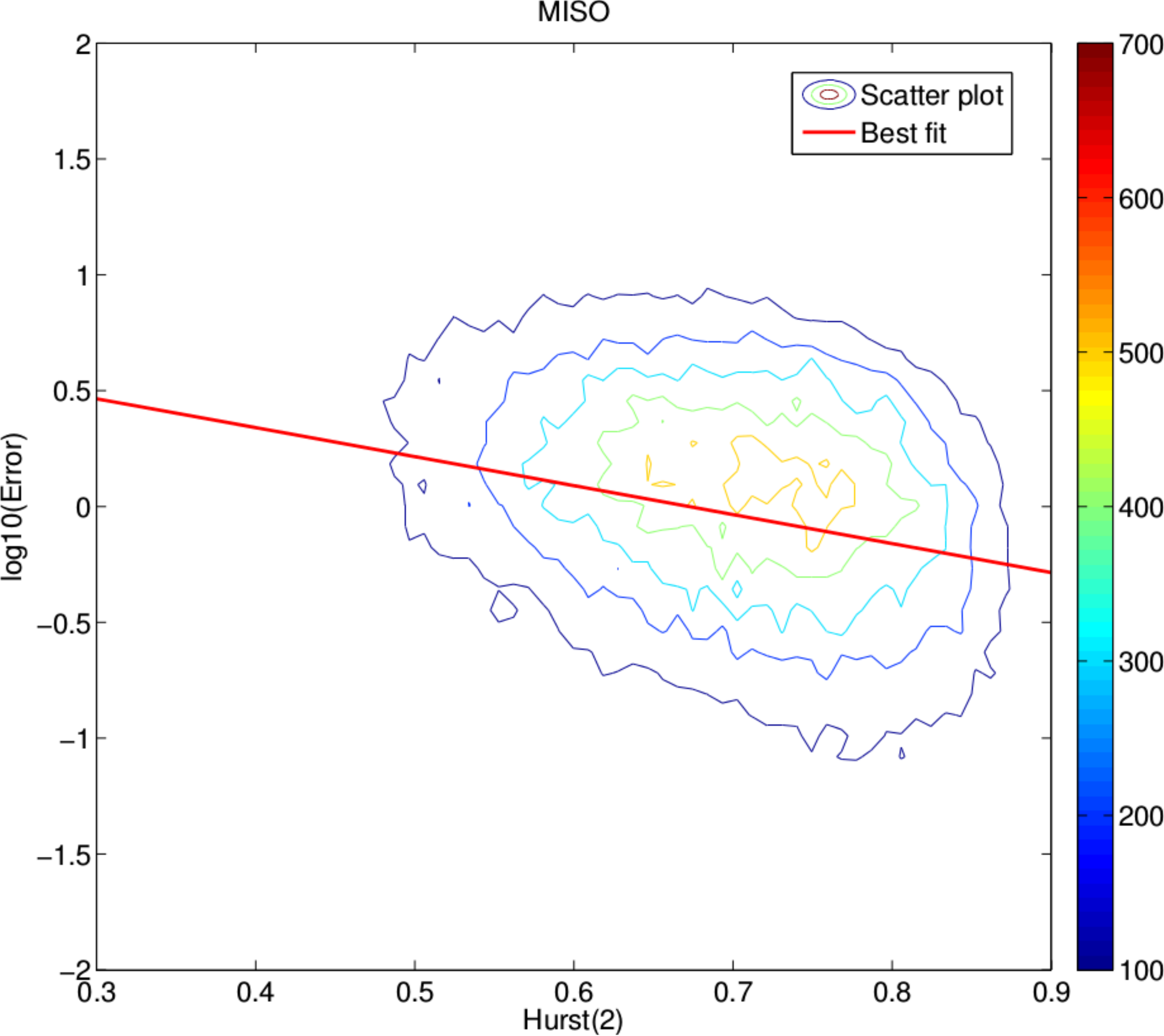}
\caption{Density plot of mean squared forecast error versus $H(q)$ for PJM (left) and MISO (right), for $q=1$ (top) and $q=2$ (bottom), for the case of a linear regression forecast and with training window $\delta h=50$. This shows that the dependence on the Hurst is negative.
The line is given by the linear fit $E(H(q)) \approx E_0 10^{c H(q)}$, with $E_0$ the intercept and $c$ is the slope in the log-linear plot. }
\label{fig:scatterploterror}
\end{figure}

We introduce the forecast error, $E_i(t,p)= |\hat S_i(t+p)-S_i(t+p)|$, with implicit training window $\delta h$.
We then compare the forecast error at time $t$ with the values of the Generalized Hurst exponents $H_i(1)$ and  $H_i(2)$ computed in the time window $[t-\delta h, t]$ for the time series $i$.
Given the prediction window $\delta h$, there are $(1632-\delta h) \times N$ errors and Generalized Hurst values, where $N$ is the number of time series in the market, and 1632 is the number of hours available for our analysis.  
%In particular, predictions at different hours ahead are analyzed in different bins. 
%Given the Hurst exponents and the prediction error values for each time series for each time, one can then evaluate the dependence of the error on the Hurst by means of a scatter plot. 
In Fig. \ref{fig:scatterploterror} we plot the density of points in the semi-log scatter plot of Generalized Hurst exponents versus forecast error for $\delta h=50$, evaluated at $p=1$ both for PJM (left) and MISO (right), together with the best fit, for both $q=1$ (top) and $q=2$ (bottom).  Similar results were obtained also for $\delta h=50 - 100$ hours. % the Hurst has been evaluated on a $\delta h=50$ hours window, the same used for the training of the linear model. 
We observe a negative dependence between the values of the Generalized Hurst exponent and the prediction error, implying that for higher Hurst exponents one has better forecasting power.
This indicate that $H(q)$ and $E(H(q))$ have a functional dependency:  $E(H(q))\approx E_0 10^{c H(q)}$ where $c$ is the slope in the log-linear fit.

\begin{figure}
\centering
\includegraphics[scale=0.38]{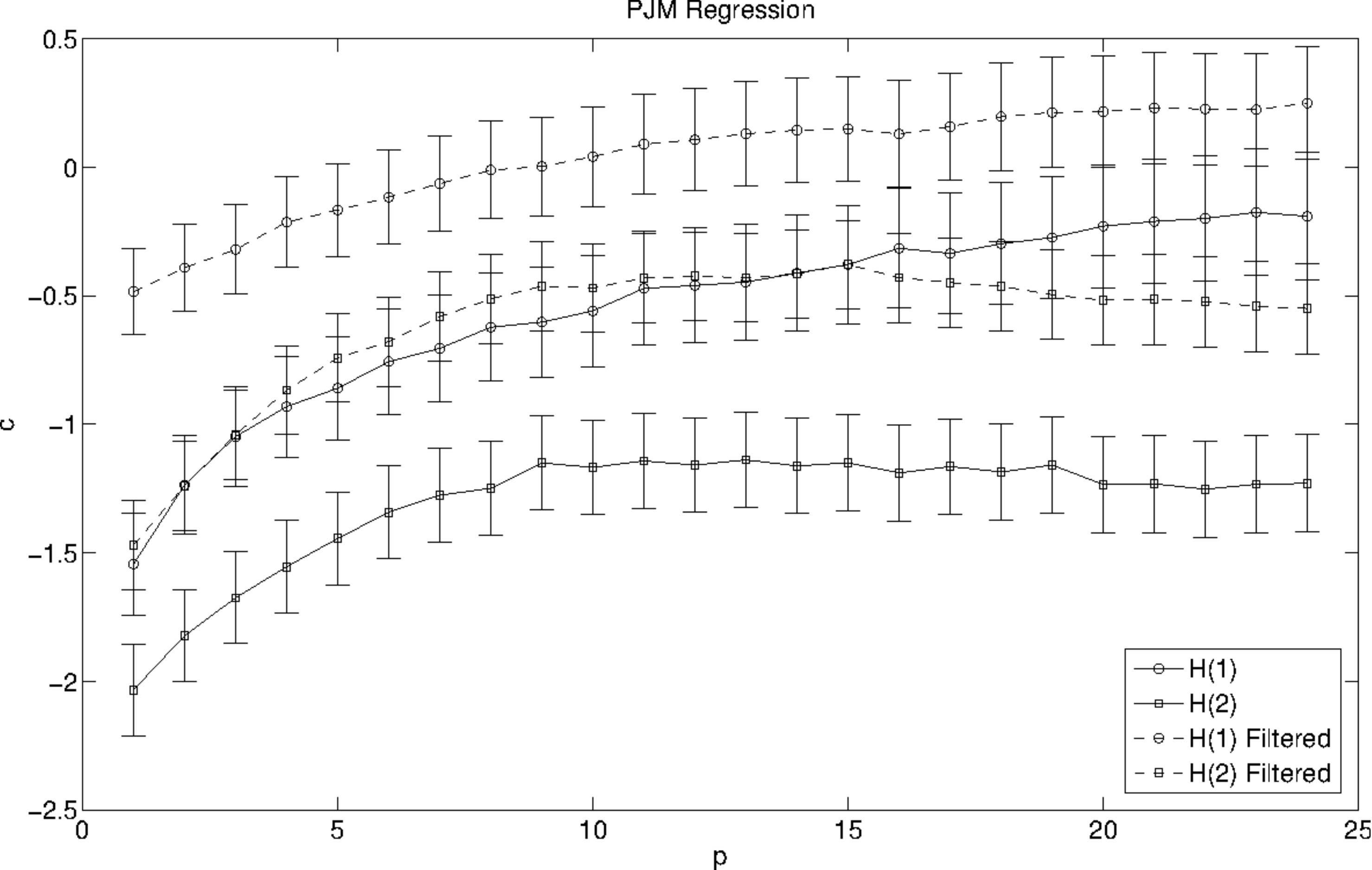}
\includegraphics[scale=0.38]{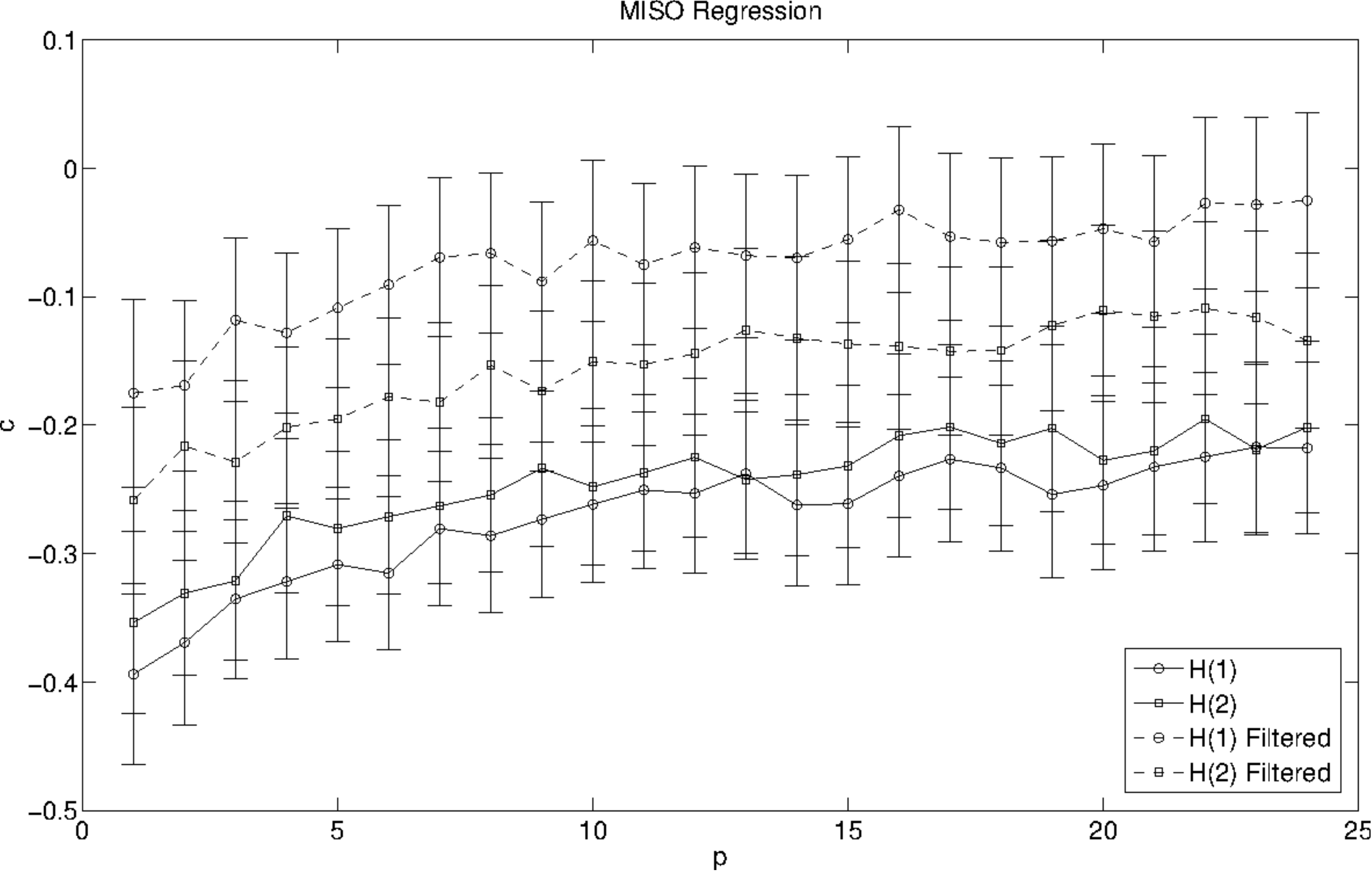}
\caption{Plot of the slopes $c$ versus the prediction lag $p$ for PJM (left) and MISO (right) as a function of $H(1)$ and $H(2)$, both for filtered (dashed line) and unfiltered (full line) signals and for $\delta h=50 $ hours.   }
\label{fig:ErrorPJMMISO}
\end{figure}

In addition, we study the effect of the seasonal components shown in Fig. \ref{fig:PowerSpectrumFFT} by taking the \textit{filtered} time series, in which the $24h$, $12h$,  $8h$ and $6h$ components have been removed. 
The results for the linear fit slope $c$ for both original and filtered time series are reported in Fig. \ref{fig:ErrorPJMMISO} (top) and Fig. \ref{fig:ErrorPJMMISO} (bottom) for $\delta h=50$ hours, for $H(1)$ and $H(2)$. 

We observe that the results are strongly affected by filtering the time series. The dashed lines (filtered) are above the full line (unfiltered), showing the importance of the strongly cyclical components for the trend of the time series as measured by the Hurst exponents. 
These results apply, qualitatively, also to the case of the MISO market, where we observe smaller values of $c$ in absolute value, but similar patterns in the curve $c(p)$ shown in Fig. \ref{fig:ErrorPJMMISO}.

This analysis suggests that the Generalized Hurst is a good estimator of trend persistence in the case of the electricity markets considered here, and we note that in general $H(2)$ is a better estimator than $H(1)$.  In addition, pries in PJM are more predictable than in MISO.

%\begin{figure}
%\centering
%\includegraphics[scale=0.3]{ScatterplotErrorMISOHurst2.jpg}
%\includegraphics[scale=0.3]{ScatterplotErrorSlopeMISOHurst2.jpg}
%\includegraphics[scale=0.3]{MISOAllH2.jpg}
%\includegraphics[scale=0.27]{MISOH1Reg.pdf} \includegraphics[scale=0.27]{MISOH1Rreg.pdf}
%\includegraphics[scale=0.27]{MISOH2Reg.pdf} \includegraphics[scale=0.27]{MISOH2Rreg.pdf}
%\includegraphics[scale=0.38]{MisoRegrH1H2.pdf}
%\caption{Plot of the slopes $c$  vs. the prediction lag $p$ for the case of MISO as a function of $H(1)$ (left) and $H(2)$ (right). In each graph, we plot the cases $\delta h=50,100,150$ both for the real (full line) and filtered signals (dashed line)}
%\label{fig:ErrorMISO}
%\end{figure}

\section{Conclusions}\label{sec:conc}

In this paper we studied the Generalized Hurst exponent for the Day-Ahead Marginal Congestion Cost components of electricity prices in two North American wholesale electricity markets. 
We observed that these prices exhibit strong multi-scaling behaviour and deviation from Brownian motion. We also observed that in the power spectrum of the time series several peaks associated to daily cycles can be identified. 
We found that the values of the Generalized Hurst exponents, $H(1)$ and $H(2)$, are are clustered between $0.7$ and $0.8$ for both markets. To our knowledge, this is the first analysis of Generalized Hurst exponents performed for these electricity markets. Doing a dynamic analysis of these exponents using moving windows of $50$ hours, we found that the generalized Hurst exponents have values which are consistently related with those evaluated on longer time windows, and have coherent market movements.

We have also shown that the Generalized Hurst exponent is a good estimator of the persistence of trends in the time series if the strongly cyclical components are taken into account, supporting the hypothesis that for higher Hurst exponents ($>\approx 0.5$) the fluctuations are trend supporting and thus simple linear models can be used to perform predictions. In general the results are different depending on the training window. In fact there is a negative correlation between the Hurst exponents and the forecast errors of the regressions, and this correlation depends on the size of the forecast horizon.  Using a training window between $50 - 100$ hours provides the best results.
In analogy to the case of stock markets, where the Hurst exponent has been connected to the efficient market hypothesis (\cite{DM2}, \cite{Lillo1}, \cite{DM1}), we have observed that indeed the Hurst exponent can be connected to the error in the prediction of returns.

\section*{Aknowledgements}
%\textcolor{red}{Tomaso Aste grants?} \textcolor{red}{Tiziana di Matteo grants?}
F.C. would like to thank Francois Lafond and Doyne J. Farmer for comments at the beginning of this work. 
The work of F.C., J.R., C. U. and A. A. was supported by Invenia Technical Computing corporation. 
TA acknowledges support of the UK Economic and Social ResearchCouncil (ESRC) in funding the Systemic Risk Centre [ES/K002309/1]. TDM wishes to thank the COST Action TD1210 for partially supporting this work.
F.C. would like to also thank LIMS for hospitality meanwhile carrying out this study.

%\textcolor{red}{thank other grants}
%\textcolor{red}{Tomaso, Tiziana add something here?}

%We have focused in particular on the relation between the generalized Hurst and the predictability of the day-ahead market by means of a simple deterministic autoregressive linear model. 

%Having observed that the generalized Hurst is not constant, we have analyzed the relation between the prediction error and the Hurst exponent, and in particular $H(1)$ and $H(2)$, as a function of the training window and for different prediction windows $n$. Both for the case of PJM and MISO, we have observed that for $n=24$ hours, there is smaller (but in some cases significant) correlation between the Hurst coefficients and prediction error.
%I




%\newpage

\end{document}